\documentclass[11pt]{article}
\usepackage{amssymb}
\usepackage{amsmath}
\usepackage{amsfonts}
 \usepackage{pifont}
\usepackage{amsmath,amsfonts,amssymb,theorem,graphicx,listings,txfonts}
\usepackage{graphics}
 \usepackage{caption2}
\usepackage{flafter}
 \usepackage{indentfirst}
\usepackage{epsfig}
  \usepackage{mathrsfs}
\usepackage{subfigure}
 \usepackage{cite}
\newtheorem{theorem}{Theorem}[section]

\newtheorem{proposition}[theorem]{Proposition}

\newtheorem{definition}[theorem]{Definition}
\newtheorem{algorithm}[theorem]{Algorithm}

\theoremstyle{example}
\newtheorem{example}[theorem]{Example}
\theoremstyle{programme}

\theoremstyle{property}
\newtheorem{property}[theorem]{Property}
\theoremstyle{problem}

\title{ Data compression of dynamic set-valued information systems}

\author
{Guangming Lang$^{a}$ \hspace{1cm} Qingguo Li$^{a}$
\thanks{Corresponding author.\quad Tel./fax: +86 731 8822855,
liqingguoli@yahoo.com.cn(G. Q. Li)
\newline\mbox{}\hspace{0.55cm}
E-mail address: langguangming1984@126.com(G. M. Lang)}\hspace{1cm}
\\
\small {$^{a}$ College of Mathematics and Econometrics, Hunan University}\\
\small {Changsha, Hunan 410082, P.R. China}}
\date{}
\begin{document}
\maketitle \baselineskip=17pt
\begin{center}
\begin{quote}
{{\bf Abstract.} This paper further investigates the set-valued
information system. First, we bring forward three tolerance
relations for set-valued information systems and explore their basic
properties in detail. Then the data compression is investigated for
attribute reductions of set-valued information systems. Afterwards,
we discuss the data compression of dynamic set-valued information
systems by utilizing the precious compression of the original
systems. Several illustrative examples are employed to show that
attribute reductions of set-valued information systems can be
simplified significantly by our proposed approach.

\\}
\end{quote}
\end{center}
\renewcommand{\thesection}{\arabic{section}}

\section{Introduction}

Rough set theory, as a powerful mathematical tool to deal with
vagueness and uncertainty of information, was proposed by
Pawlak\cite{Pawlak1,Pawlak2,Pawlak3,Pawlak4} in the early 1980s. But
the requirement of the equivalence relation limits the applications
of rough sets in many practical situations. To apply rough set
theory to more complex data sets, it has been extended by combining
with fuzzy
sets\cite{He1,Banerjee1,Bhatt1,Biswas1,Bobillo1,Capotorti1,Chakrabarty1,
Dubois1,Jensen1,Morsi1,Nanda1,Feng2}, probability
theory\cite{Slezak1,Skowron1,Yao1,Yao2,Yao3,Ziarko1,Dai1},
topology\cite{Diker1,Feng1,Yang1,Zakowski1,Zhu1,Yang2} and matroid
theory\cite{Wang2}.

Originally, the theory of rough sets based data analysis starts from
the single-valued information system. In practice, it may often
happen that some of attribute values for an object are set-valued.
Recently, the set-valued information system has become a rapidly
developing research area and got a lot of attention. For example,
Guan et al.\cite{Guan1} initially introduced the set-valued
information system as generalized models of single-valued
information systems. Then Qian et al.\cite{Qian1} studied the
set-valued ordered information system. Afterwards, many
researchers\cite{Liu1,Liu2,Liu3,Li1,Chen1,Zhang1} investigated the
dynamic set-valued information system. In the literature
\cite{Guan1}, the tolerance relation which discerns objects on the
basis of that whether there exists common attribute values or not
neglects some other difference. For example, it may happen that
there are two (respectively, ten) common values between objects A
and B (respectively, A and C) with respect to an attribute, and
objects B and C belong to the same tolerance class of object A.
Although the number of common attribute values between objects A and
B is larger than that between objects A and C, the tolerance
relation cannot discern objects B and C in the tolerance class of
object A. Therefore, it is of interest to introduce some tolerance
relations for solving the above issue.

Meanwhile,
homomorphisms\cite{Grzymala-Busse2,Li2,Gong1,Wang1,Zhu2,Zhu3,Liu2}
have been considered as an important approach for attribute
reductions of information systems. For instance,
Grzymala-Busse\cite{Grzymala-Busse2} initially introduced seven
kinds of homomorphisms of knowledge representation systems and
investigated their basic properties in detail. Then Li et
al.\cite{Li2} investigated invariant characters of information
systems under some homomorphisms. Afterwards, many
scholars\cite{Gong1,Wang1,Zhu2,Zhu3} discussed the relationship
between information systems by means of homomorphisms. In practical
situations, there exist a great many set-valued information systems.
Inspired by the above work, attribute reductions of set-valued
information systems may be conducted by means of homomorphisms. But
so far few attempts have been made on the data compression of
set-valued information systems under the condition of homomorphisms.
In addition, the information system varies with time due to the
dynamic characteristics of data collection, and the non-incremental
approach to compressing the dynamic set-valued information system is
often very costly or even intractable. Therefore, it is interesting
to apply an incremental updating scheme to maintain the compression
dynamically and avoid unnecessary computations by utilizing the
compression of the original set-valued information system.

The purpose of this paper is to study the set-valued information
system further. First, we introduce three tolerance relations for
the set-valued information system and investigate their basic
properties. Subsequently, the discernibility matrix based on the
proposed relation is presented for attribute reductions of
set-valued information systems. Second, we discuss the data
compression of set-valued information systems. Concretely, a
large-scale set-valued information system can be compressed into a
relative-small relation information system under the condition of a
homomorphism, and their attribute reductions are equivalent to each
other.  Third, the data compression of dynamic set-valued
information systems is investigated by utilizing the precious
compression of the original information systems. There are four
types of dynamic set-valued information systems: adding and deleting
attributes, adding and deleting objects. Using the proposed
approach, the time complexity for computing attribute reducts of
set-valued information systems can be reduced greatly by avoiding
unnecessary computations.

The rest of this paper is organized as follows: Section 2 briefly
reviews the basic concepts of set-valued information systems and
consistent functions. In Section 3, we put forward three tolerance
relations for the set-valued information system and investigate
their basic properties in detail. We also present the discernibility
matrix based on the proposed relation. Section 4 is devoted to
discussing the data compression of set-valued information systems.
In Section 5, we investigate the data compression of dynamic
set-valued information systems. We conclude the paper in Section 6.

\section{Preliminaries}

In this section, we briefly review some concepts of the set-valued
information system and the relation information system. In addition,
an example is employed to illustrate the set-valued information
system.

\begin{definition}\cite{Guan1}
Suppose $S=(U, A, V, f)$ is a set-valued information system, where
$U=\{x_{1},x_{2},...,x_{n}\}$ is a non-empty finite set of objects,
$A=\{a_{1},a_{2},...,a_{m}\}$ is a non-empty finite set of
attributes,V is the set of attribute values, f is a mapping from
$U\times A$ to $ V$, where $f: U\times A \longrightarrow 2^{V}$ is a
set-valued mapping.
\end{definition}

It is obvious that the classical information system can be regarded
as a special case of the set-valued information system. There are
many semantic interpretations for the set-valued information system,
we summarize two types of them as follows:

Type 1: For $x\in U, a\in A$, $f(x,a)$ is interpreted conjunctively.
For example, if $a$ is the attribute ``speaking language", then
$f(x,a)$=$\{$German, French, Polish$\}$ can be viewed as: $x$ speaks
German, French and Polish, and $x$ can speak three languages.

Type 2: For $x\in U, a\in A$, $f(x,a)$ is interpreted disjunctively.
For instance, if $a$ is the attribute ``speaking language", then
$f(x,a)$=$\{$German, French, Polish$\}$ can be regarded as: $x$
speaks German, French or Polish, and $x$ can speak only one of them.

\begin{definition}\cite{Guan1}
Let $S=(U, A, V, f)$ be a set-valued information system, $a\in A$,
and $B\subseteq A$. Then the tolerance relations $R_{a}$ and $R_{B}$
are defined as
\begin{eqnarray*}
R_{a}&=&\{(x,y)| f(x,a)\cap f(y,a)\neq \emptyset, x,y\in U\};\\
R_{B}&=&\{(x,y)| \forall b\in B, f(x,b)\cap f(y,b)\neq \emptyset,
x,y\in U\}.
\end{eqnarray*}
\end{definition}

In other words, $(x,y)\in R_{B}$ is viewed as $x$ and $y$ are
indiscernible with respect to $B$, and $R_{B}(x)$ is seen as the
tolerance class for $x$ with respect to $B$. Naturally,
$R_{B}=\bigcap_{b\in B}R_{b}$. In spite of that the tolerance
relation has been applied successfully in many fields, there exist
some issues which need to be solved in practical situations. We
employ an example to illustrate the problems of the tolerance
relation presented in Definition 2.2 as below.

\begin{table}[htbp]
\caption{A set-valued information system.}
 \tabcolsep0.47in
\begin{tabular}{c c c c c}
\hline $U$  & $a_{1}$ &$a_{2}$& $a_{3}$& $a_{4}$\\ \hline
$x_{1}$ & $\{0\}$& $\{0\}$ &$\{1,2\}$& $\{1,2\}$ \\
$x_{2}$ & $\{0,1,2\}$& $\{1,2\}$  &$\{1,2\}$& $\{0,1,2\}$\\
$x_{3}$ & $\{1,2\}$& $\{1\}$  &$\{1\}$& $\{1,2\}$\\
$x_{4}$ & $\{0,1\}$& $\{0,2\}$  &$\{1,2\}$& $\{1,2\}$\\
$x_{5}$ & $\{1,2\}$& $\{1,2\}$ &$\{1,2\}$  & $\{1\}$ \\
$x_{6}$ & $\{1\}$& $\{1\}$&$\{0,1\}$ & $\{0,1\}$\\
\hline
\end{tabular}
\end{table}

\begin{example} Table 1 depicts a set-valued information system.
In the sense of Definition 2.2, $R_{a_{1}}(x_{2})=\{x_{1}, x_{2},
x_{3}, x_{4}, x_{5}, x_{6}\}$. Obviously, we have that $(x_{1},
x_{2}), (x_{3}, x_{2})\in R_{a_{1}}$. But $|f(x_{1}, a_{1})\cap
f(x_{2}, a_{1})|=1$ and $|f(x_{2}, a_{1})\cap f(x_{3}, a_{1})|=2$.
Furthermore, we obtain that $(x_{1},x_{4}), (x_{6},x_{4})\in
R_{a_{1}}.$ But $f(x_{1}, a_{1})\cap f(x_{4}, a_{1})=\{0\}$ and
$f(x_{6}, a_{1})\cap f(x_{4}, a_{1})=\{1\}$. Although there are some
difference between objects which are in the same tolerance class,
$R_{a_{1}}$ cannot discern them.
\end{example}

To compress the relation information system, Wang et al. presented
the concept of consistent functions as follows.

\begin{definition}\cite{Wang1}
Let $U_{1}$ and $U_{2}$ be two universes, $f$ a mapping from $U_{1}$
to $U_{2}$, the relation $R$ a mapping from $ U\times U$ to $\{0,
1\}$, and $[x]_{f}=\{y\in U_{1}|f(x)=f(y)\}$. For any $x,y\in
U_{1}$, if $R(u,v)=R(s,t)$ for any two pairs $(u,v),(s,t)\in
[x]_{f}\times[y]_{f}$, then $f$ is said to be consistent with
respect to $R$.
\end{definition}

Especially, if the consistent function is a surjection, then it is a
homomorphism between relation information systems. We can compress a
large-scale information system into a relatively small-scale one
under the condition of a homomorphism. It has been proved that
attribute reductions of the original system and image system are
equivalent to each other. Therefore, the consistent functions
provide an approach to studying the data compression of relation
information systems.

\section{ The tolerance relation based the discernibility matrix
for set-valued information systems}

In this section, we propose three tolerance relations to address the
problem illustrated in Example 2.3. Then we present the concept of a
discernibility matrix based on the proposed tolerance relation for
attribute reductions of set-valued information systems.

\begin{definition}
Let (U, A, V, f) be a set-valued information system, $a\in A$, and
$B\subseteq A$. Then the tolerance relations $R^{\geq h}_{a}$ and
$R^{\geq I_{B}}_{B}$ are defined as
\begin{eqnarray*}
R^{\geq h}_{a}&=&\{(x, y)||f(x,a)\cap f(y,a)|\geq h, x, y\in U\};\\
R^{\geq H_{B}}_{B}&=&\{(x, y)| |f(x,a_{i})\cap f(y,a_{i})| \geq
h_{i}, x, y\in U, a_{i}\in B\},
\end{eqnarray*}
where $|\cdot|$ denotes the cardinality of a set, $H_{B}=(h_{1},
h_{2},...,h_{m})$ and $h_{i}=0$ if $a_{i}\notin B$.
\end{definition}

From Definition 3.1, we see that the number of common attribute
values between objects are considered in the tolerance relations.
Furthermore, we obtain that $R_{a}=R^{\geq 1}_{a}$, $R^{\geq
(1,1,...,1)}_{B}=R_{B}$ and $R^{\geq H_{B}}_{B}=\bigcap_{a_{i}\in B}
R^{\geq h_{i}}_{a_{i}}$. For the convenient representation, we
denote $R^{\geq H_{B}}_{B}(x)=[x]^{\geq H_{B}}_{B}=\{y|(x,y)\in
R^{\geq H_{B}}_{B}\}$ in the following. We define that $K=(k_{1},
k_{2},...,k_{m})\leq H_{B}$ if and only if $k_{i}\leq h_{i}$ for
$1\leq i\leq m$. Specially, if $\{R^{\geq h}_{a}(x)|x\in U\}$ is a
covering of $U$, then $R^{\geq h}_{a}$ is called the $\geq
h-$relation. In general,  $R^{\geq h}_{a}$ and $R^{\geq H_{B}}_{B}$
are symmetric and intransitive, $R^{\geq h}_{a}$ and $R^{\geq
H_{B}}_{B}$ are not reflexive necessarily if $h>1$ and
$H_{B}\neq(1,1,...,1)$, respectively. For example, consider Table 1,
we obtain that $R^{\geq 2}_{a_{1}}(x_{1})=\emptyset$. That is,
$(x_{1},x_{1})\notin R^{\geq 2}_{a_{1}}.$

\begin{proposition}
Let (U, A, V, f) be a set-valued information system, and $B,
C\subseteq A$. Then we have

$(1)$ if $H_{B}\leq H_{C}\leq H_{A}$, then $R^{\geq
H_{A}}_{A}\subseteq R^{\geq H_{C}}_{C}\subseteq R^{\geq H_{B}}_{B}$;

$(2)$ if $H_{B}\leq H_{C}\leq H_{A}$, then $[x]^{\geq
H_{A}}_{A}\subseteq [x]^{\geq H_{C}}_{C}\subseteq [x]^{\geq
H_{B}}_{B}$.
\end{proposition}

We notice that $[y]^{\geq H_{B}}_{B}\subseteq [x]^{\geq H_{B}}_{B}$
does not hold necessarily if $y\in [x]^{\geq H_{B}}_{B}$, and that
$[y]^{\geq H_{B}}_{B}=[x]^{\geq H_{B}}_{B}$ does not imply $x=y$,
which can be illustrated by the following example.

\begin{example}
Consider Table 1, we obtain that $[x_{1}]^{\geq
(1,0,0,0)}_{\{a_{1}\}}=\{x_{1}, x_{2}, x_{4}\}.$ It is clear that
$x_{2}\in [x_{1}]^{\geq (1,0,0,0)}_{\{a_{1}\}}$ and $[x_{2}]^{\geq
(1,0,0,0)}_{\{a_{1}\}}=\{x_{1}, x_{2}, x_{3}, x_{4}, x_{5},
x_{6}\}$. Moreover, we have that $[x_{4}]^{\geq
(1,0,0,0)}_{\{a_{1}\}}=\{x_{1}, x_{2}, x_{3}, x_{4}, x_{5},
x_{6}\}$. Thus $[x_{2}]^{\geq (1,0,0,0)}_{\{a_{1}\}}=[x_{4}]^{\geq
(1,0,0,0)}_{\{a_{1}\}}$. But $x_{2}\neq x_{4}$.
\end{example}

\begin{definition}
Let $S=(U, A, V, f)$ be a set-valued information system,
$\mathscr{R}^{\geq}_{A}=\{R^{\geq _{h_{1}}}_{a_{1}}, R^{\geq
_{h_{2}}}_{a_{2}},...,$ $ R^{\geq _{h_{m}}}_{a_{m}}\}$, and $R^{\geq
_{h_{i}}}_{a_{i}}$ the $\geq h_{i}-$relation. Then $(U,
\mathscr{R}^{\geq}_{A})$ is called the induced $\geq-$relation
information system of S.
\end{definition}

For the sake of convenience, we denote $R^{\geq _{h_{i}}}_{a_{i}}$
as $R_{i}$ and consider the situation that $h_{i}=1$ in the
following. An example is employed to illustrate the induced
$\geq-$relation information system.

\begin{example}
Consider Table 1, we obtain the induced $\geq-$relation information
system $(U, \mathscr{R}^{\geq}_{A})$ and
$\mathscr{R}^{\geq}_{A}=\{R_{i}|1\leq i\leq 4\}$, where
\begin{eqnarray*}
R_{1}(x_{1})&=&\{x_{1}, x_{2}, x_{4}\},
R_{1}(x_{2})=R_{1}(x_{4})=\{x_{1},x_{2},x_{3},x_{4},x_{5},x_{6}\},
R_{1}(x_{3})=R_{1}(x_{5})=R_{1}(x_{6})=\{x_{2},x_{3},x_{4},\\&&x_{5},x_{6}\};\\
R_{2}(x_{1})&=&\{x_{1},x_{4}\},
R_{2}(x_{2})=R_{2}(x_{5})=\{x_{2},x_{3},x_{4},x_{5},x_{6}\},
R_{2}(x_{3})=R_{2}(x_{6})=\{x_{2},x_{3},x_{5},x_{6}\},
R_{2}(x_{4})=\{x_{1},\\&&x_{2},x_{4},x_{5}\};
\\
R_{3}(x_{1})&=&R_{3}(x_{2})=R_{3}(x_{3})=R_{3}(x_{4})
=R_{3}(x_{5})=R_{3}(x_{6})=\{x_{1},x_{2},x_{3},x_{4},x_{5},x_{6}\};\\
R_{4}(x_{1})&=&R_{4}(x_{2})=R_{4}(x_{3})=R_{4}(x_{4})
=R_{4}(x_{5})=R_{4}(x_{6})=\{x_{1},x_{2},x_{3},x_{4},x_{5},x_{6}\}.
\end{eqnarray*}
\end{example}

\begin{definition}
Let $S=(U, A, V, f)$ be a set-valued information system, $(U,
\mathscr{R}^{\geq}_{A})$ the induced $\geq-$relation information
system of $S$, and $P\subseteq A$. If
$\bigcap\mathscr{R}^{\geq}_{P}=\bigcap\mathscr{R}^{\geq}_{A}$ and
$\bigcap\mathscr{R}^{\geq}_{P^{\ast}}\neq\bigcap\mathscr{R}^{\geq}_{A}$
for any $\mathscr{R}^{\geq}_{P^{\ast}}\subsetneqq
\mathscr{R}^{\geq}_{P}$, then $\mathscr{R}^{\geq}_{P}$ is called a
reduct of $(U, \mathscr{R}^{\geq}_{A})$.
\end{definition}

By Definition 3.6, we see that the reduct is the minimal subset of
attribute set, which preserves the relation
$\mathscr{R}^{\geq}_{A}$. For instance, we get the reduct
$P=\{R_{2}\}$ in the sense of Definition 3.6 for the relation
information system presented in Example 3.5.

Now we introduce the discernibility matrix based on Definition 3.1
and investigate its basic properties.

\begin{definition}
Let $S=(U, A, V, f)$ be a set-valued information system. Then its
discernibility matrix $M_{A}=(M(x,y))$ is a $|U|\times|U|$ matrix,
the element $M(x,y)$ is defined by \makeatother $$M(x,y)=\{a\in
A|(x,y)\notin R^{\geq h_{a}}_{a}, x, y\in U\},$$ where $R^{\geq
h_{a}}_{a}$ is a $\geq h_{a}-$relation.
\end{definition}

That is, the physical meaning of the matrix element $M(x,y)$ is that
objects $x$ and $y$ can be distinguished by any element of $M(x,y)$.
If we obtain that $M(x,y)\neq \emptyset$, then objects $x$ and $y$
can be discerned. It is sufficient to consider only the lower
triangle or the upper triangle of the matrix since the
discernibility matrix $M$ is symmetric.

\begin{definition}
Let $S=(U, A, V, f)$ be a set-valued information system, and
$M=(M(x,y))$ the discernibility matrix of $S$. Then
$\triangle=\bigwedge_{(x,y)\in U^{2}}\bigvee M(x,y)$ is called the
discernibility function of $S$.
\end{definition}

The expression $\bigvee M(x,y)$ denotes the disjunction of all
attributes in $M(x,y)$, and the expression $\bigwedge\{\bigvee
M(x,y)\}$ stands for the conjunction of all $\bigvee M(x,y)$. In
addition, $\bigwedge B$ is a prime implicant of the discernibility
function $\triangle$ if and only if $B$ is a reduct of $S$.

Next, we propose another two concepts of tolerance relations and
discuss their basic properties for set-valued information systems.

\begin{definition}
Let (U, A, V, f) be a set-valued information system, $a\in A$, and
$B\subseteq A$. Then the tolerance relations $R^{h}_{a}$ and
$R^{H_{B}}_{B}$ are defined as
\begin{eqnarray*}
R^{ h}_{a}&=&\{(x, y)||f(x,a)\cap f(y,a)|=h, x,
y\in U\};\\
R^{H_{B}}_{B}&=&\{(x, y)||f(x,a_{i})\cap f(y,a_{i})| =h_{i}, x, y\in
U, a_{i} \in B\}.
\end{eqnarray*}

\end{definition}

From Definition 3.9, we see that $R^{h}_{a}$ and $R^{H_{B}}_{B}$ are
symmetric and intransitive, $R^{h}_{a}$ and $R^{H_{B}}_{B}$ are not
reflexive necessarily. Meanwhile, we have that $R^{\geq
h}_{a}=\bigcup_{j\geq h} R^{j}_{a}$ and $R^{\geq
H_{B}}_{B}=\bigcup_{K\geq H_{B}}R^{K}_{B}$. For the sake of
simplicity, we note that
$R^{H_{B}}_{B}(x)=[x]^{H_{B}}_{B}=\{y|(x,y)\in R^{H_{B}}_{B}\}$.

\begin{property}
Let (U, A, V, f) be a set-valued information system, and $B,
C\subseteq A$. Then we have

$(1)$ if $H_{B}\leq H_{C}\leq H_{A}$, then $R^{H_{A}}_{A}\subseteq
R^{ H_{C}}_{C}\subseteq R^{H_{B}}_{B}$;

$(2)$ if $H_{B}\leq H_{C}\leq H_{A}$, then $[x]^{H_{A}}_{A}\subseteq
[x]^{H_{C}}_{C}\subseteq [x]^{H_{B}}_{B}$.
\end{property}

\begin{definition}
Let (U, A, V, f) be a set-valued information system, $a\in A$,
$B\subseteq A$, and $P\subseteq V_{a}$. Then the tolerance relations
$R^{P}_{a}$ and $R^{\mathscr{P}}_{B}$ are defined as
\begin{eqnarray*}
R^{P}_{a}&=&\{(x, y)|f(x,a)\cap f(y,a)=P, x, y\in U\};\\
R^{\mathscr{P}}_{B}&=&\{(x, y)| f(x,a_{i})\cap f(y,a_{i}) =P_{i}, x,
y\in U, a_{i} \in B\},
\end{eqnarray*}
where $\mathscr{P}=(P_{1}, P_{2},...,P_{m})$, and $P_{i}$ is defined
as $P_{i}\subseteq V_{a_{i}}$ $(respectively, P_{i}=\emptyset)$ if
$a_{i}\in B$ $(respectively, a_{i}\notin B)$.
\end{definition}

In the sense of Definitions 3.9 and 3.11, it is observed that
$R^{h}_{a}=\bigcup\{R^{P}_{a}|P\in 2^{A}, |P|=h\}$. Furthermore,
$R^{P}_{a}$ and $R^{\mathscr{P}}_{B}$ are symmetric and
intransitive. By Definitions 3.1, 3.9 and 3.11, we obtain that
$$R^{\geq h}_{a}=\bigcup_{i\geq h} R^{i}_{a}=\bigcup_{i\geq h}
\bigcup\{R^{P}||P|=i, P\in 2^{V_{a}}\}$$ and
$$R^{\geq I}_{B}=\bigcap_{a\in B}\{\bigcup_{i\geq h} R^{i}_{a}\}=
\bigcap_{a\in B}\{\bigcup_{i\geq h} \bigcup\{R^{P}||P|=i, P\in
2^{V_{a}}\}\}.$$ In addition, we can define discernibility matrixes
based on Definitions 3.9 and 3.11, respectively. For the sake of
simplicity, we do not present them in this section.

\section{Data compression of the set-valued information system}

In this section, we investigate the data compression of the
large-scale set-valued information system. Concretely, we derive the
induced $\geq-$relation information system of the set-valued
information system. Then the induced $\geq-$relation information
system is compressed into a relatively small one under the condition
of a homomorphism, and attribute reductions of the original system
and image system are equivalent to each other. In addition, we
illustrate that the time complexity of computing attribute
reductions can be reduced greatly by means of the compression from
another view.

\begin{definition}
Let $(U_{1}, \mathscr{R}^{\geq}_{A})$ be the induced $\geq-$relation
information system of the set-valued information system $S=(U_{1},
A, V, f)$, $R\in \mathscr{R}^{\geq}_{A}$, $[x]_{R}=\{y|R(x)=R(y),
x,y\in U_{1}\}$, and $U_{1}/R=\{[x]_{R}|x\in U_{1}\}$. Then
$U_{1}/R$ is called the partition based on $R$.
\end{definition}

Following, we employ Table 2 to show the partition based on each
relation for the induced $\geq-$relation information system $(U_{1},
\mathscr{R}^{\geq}_{A})$, where $P_{ix_{j}}$ stands for the block
containing $x_{j}$ in the partition based on the relation $R_{i}$.
It is easy to see that $P_{A x_{j}}=\bigcap_{1\leq i\leq
m}P_{ix_{j}}$, where $P_{A x_{j}}$ denotes the block containing
$x_{j}$ in the partition based on $\mathscr{R}^{\geq}_{A}$.
\begin{table}[htbp]
\caption{The partitions based on each relation $R_{i}$ $(1\leq i\leq
m )$ and $\mathscr{R}^{\geq}_{A}$, respectively.}
 \tabcolsep0.31in
\begin{tabular}{c c c c c c c c}
\hline $U_{1}$  &$R_{1}$& $R_{2}$&.&.&. &$R_{m}$& $\mathscr{R}^{\geq}_{A}$\\
\hline
$x_{1}$ & $P_{1x_{1}}$& $P_{2x_{1}}$&.&.&. &$P_{mx_{1}}  $ & $P_{A x_{1}}$\\
$x_{2}$ & $P_{1x_{2}}$& $P_{2x_{2}}$&.&.&. &$P_{mx_{2}}  $&$P_{A x_{2}}$\\
$.$     & $.$        & .&.&.&. &$ . $& .\\
$.$     & $.$        & .&.&.&. &$ . $&.\\
$.$     & $.$        & .&.&.&. &$ . $ &.\\
$x_{n}$ & $P_{1x_{n}}$& $P_{2x_{2}}$&.&.&. &$P_{mx_{n}}$&$P_{A x_{n}}$  \\
\hline
\end{tabular}
\end{table}

We present the algorithm of compressing the set-valued information
system as follows.

\begin{algorithm}
Let $S=(U_{1}, A, V, f)$ be a set-valued information system, where
$U_{1}=\{x_{1},...,x_{n}\}$ and $A=\{a_{1},...,a_{m}\}$.

Step 1. Input the set-valued information system $S=(U_{1},A,V,f)$
and obtain the induced $\geq-$relation information system
$(U_{1},\mathscr{R}^{\geq}_{A})$, where
$\mathscr{R}^{\geq}_{A}=\{R_{1}, R_{2}, ...,R_{m}\}$;

Step 2. Compute the partition $U_{1}/R_{i}$ $(1\leq i\leq m )$ and
obtain $U_{1}/\mathscr{R}^{\geq}_{A}=\{C_{i}|1\leq i\leq N\}$;

Step 3. Define the function $g(x)=y_{i}$ for any $x\in C_{i}$ and
obtain $(U_{2}, g(\mathscr{R}^{\geq}_{A}))$, where
$U_{2}=\{g(x_{i})|x_{i}\in U_{1}\}$ and
$g(\mathscr{R}^{\geq}_{A})$=$\{g(R_{1}), g(R_{2}),...,g(R_{m})\}$;

Step 4. Obtain attribute reductions $\{g(R_{i1}), g(R_{i2}),
...,g(R_{ik})\}$ of $(U_{2}, \{g(R_{1}), g(R_{2}), ...,g(R_{m})\})$;

Step 5. Obtain a reduct $\{R_{i1}, R_{i2}, ...,R_{ik}\}$ of $(U_{1},
\mathscr{R}^{\geq}_{A})$ and output the results.
\end{algorithm}

The mapping $g$ presented in Algorithm 4.2 is a homomorphism from
$(U_{1},\mathscr{R}^{\geq}_{A})$ to
$(U_{2},g(\mathscr{R}^{\geq}_{A}))$ in the sense of Definition 2.4,
and attribute reductions of $(U_{1},\mathscr{R}^{\geq}_{A})$ and
$(U_{2},g(\mathscr{R}^{\geq}_{A}))$ are equivalent to each other
under the condition of the homomorphism $g$.

\noindent\textbf{Remark.} In Example 3.1\cite{Wang1}, Wang et al.
only obtained the partition $U_{1}/\mathscr{R}^{\geq}_{A}$. But we
get $U_{1}/\mathscr{R}^{\geq}_{A}$ by computing $U_{1}/R_{i}$ for
any $R_{i}\in \mathscr{R}^{\geq}_{A}$ in Algorithm 4.2. By using the
proposed approach, the data compression of dynamic set-valued
information systems can be conducted on the basis of that of the
original set-valued information system, which is illustrated in
Section 5.

We give an example to show the data compression of set-valued
information systems with Algorithm 4.2.

\begin{table}[htbp]
\caption{A set-valued information system.}
 \tabcolsep0.45in
\begin{tabular}{c c c c c}
\hline $U_{1}$  & $a_{1}$ &$a_{2}$& $a_{3}$& $a_{4}$\\ \hline
$x_{1}$ & $\{0\}$& $\{0\}$ &$\{1,2\}$& $\{1,2\}$ \\
$x_{2}$ & $\{0,1,2\}$& $\{0,1,2\}$  &$\{1,2\}$& $\{0,1,2\}$\\
$x_{3}$ & $\{1,2\}$& $\{0,1\}$  &$\{1,2\}$& $\{1,2\}$\\
$x_{4}$ & $\{0,1\}$& $\{0,2\}$  &$\{1,2\}$& $\{1\}$\\
$x_{5}$ & $\{1,2\}$& $\{1,2\}$ &$\{1,2\}$  & $\{1\}$ \\
$x_{6}$ & $\{1\}$& $\{1,2\}$&$\{0,1\}$ & $\{0,1\}$\\
$x_{7}$ & $\{0\}$& $\{0\}$ &$\{1,2\}$& $\{1,2\}$ \\
$x_{8}$ & $\{1\}$& $\{1,2\}$&$\{0,1\}$ & $\{0,1\}$\\
\hline
\end{tabular}
\end{table}

\begin{example} Table 3 depicts the set-valued information system $S_{1}=(U_{1}, A, V, f)$.
According to Definitions 3.1 and 3.4, we obtain the induced
$\geq-$relation information system $(U_{1},
\mathscr{R}^{\geq}_{A})$, and
$\mathscr{R}^{\geq}_{A}=\{R_{1},R_{2},R_{3},R_{4}\},$ where
\begin{eqnarray*}
R_{1}(x_{1})&=&R_{1}(x_{7})=\{x_{1}, x_{2}, x_{4}, x_{7}\},
R_{1}(x_{2})=R_{1}(x_{4})=\{x_{1},x_{2},x_{3},x_{4},x_{5},x_{6},
x_{7}, x_{8}\},\\
R_{1}(x_{3})&=&R_{1}(x_{5})=R_{1}(x_{6})=R_{1}(x_{8})=\{x_{2},x_{3},x_{4},x_{5},x_{6},x_{8}\};
\\ R_{2}(x_{1})&=&R_{1}(x_{7})=\{x_{1}, x_{2},x_{3}, x_{4}, x_{7}\},
R_{2}(x_{2})=R_{2}(x_{3})=R_{2}(x_{4})=\{x_{1},x_{2},x_{3},x_{4},x_{5},x_{6},x_{7},x_{8}\},\\
R_{2}(x_{5})
&=&R_{2}(x_{6})=R_{2}(x_{8})=\{x_{2},x_{3},x_{4},x_{5},x_{6},x_{8}\};
\\
R_{3}(x_{1})&=&R_{3}(x_{2})=R_{3}(x_{3})=R_{3}(x_{4})
=R_{3}(x_{5})=R_{3}(x_{6})=R_{3}(x_{7})=R_{3}(x_{8})=\{x_{1},x_{2},x_{3},x_{4},x_{5},x_{6},x_{7},x_{8}\};\\
R_{4}(x_{1})&=&R_{4}(x_{2})=R_{4}(x_{3})=R_{4}(x_{4})
=R_{4}(x_{5})=R_{4}(x_{6})=R_{4}(x_{7})=R_{4}(x_{8})=\{x_{1},x_{2},x_{3},x_{4},x_{5},x_{6},x_{7},x_{8}\}.
\end{eqnarray*}

For the sake of convenience, we present $\{R_{i}(x_{j})|x_{j}\in
U_{1}\}$ instead of $R_{i}$ in this work. By Definition 4.1, we
derive the partitions $U_{1}/R_{1}$, $U_{1}/R_{2}$, $U_{1}/R_{3}$
and $U_{1}/R_{4}$ shown in Table 4. Then, based on $U_{1}/R_{1}$,
$U_{1}/R_{2}$, $U_{1}/R_{3}$ and $U_{1}/R_{4}$, we get the partition
$U_{1}/\mathscr{R}^{\geq}_{A}=\{\{x_{1},x_{7}\},\{x_{2},x_{4}\},
\{x_{3}\},\{x_{5},x_{6},x_{8}\}\}$ and define a mapping $g:
U_{1}\longrightarrow U_{2}$ as follows:
$$g(x_{1})=g(x_{7})= y_{1}, g(x_{2}) = g(x_{4}) = y_{2},
g(x_{3})= y_{3}, g(x_{5})=g(x_{6})=g(x_{8})= y_{4}.$$
Afterwards, we
derive the compressed relation information system
$(U_{2},g(\mathscr{R}^{\geq}_{A}))$, where
$U_{2}=\{y_{1},y_{2},y_{3},y_{4}\}$,
$g(\mathscr{R}^{\geq}_{A})=\{g(R_{1}), g(R_{2}), g(R_{3}),
g(R_{4})\}$, and
\begin{eqnarray*}
g(R_{1})(y_{1})&=&\{y_{1}, y_{2}\},
g(R_{1})(y_{2})=\{y_{1},y_{2},y_{3},y_{4}\},
g(R_{1})(y_{3})=g(R_{1})(y_{4})=\{y_{2},y_{3},y_{4}\};\\
g(R_{2})(y_{1})&=&\{y_{1}, y_{2},y_{3}\},
g(R_{2})(y_{2})=g(R_{2})(y_{3})=\{y_{1},y_{2},y_{3},y_{4}\},
g(R_{2})(y_{4})=\{y_{2},y_{3},y_{4}\};\\
g(R_{3})(y_{1})&=&g(R_{3})(y_{2})=g(R_{3})(y_{3})=g(R_{3})(y_{4})
=\{y_{1},y_{2},y_{3},y_{4}\};\\
g(R_{4})(y_{1})&=&g(R_{4})(y_{2})=g(R_{4})(y_{3})=g(R_{4})(y_{4})
=\{y_{1},y_{2},y_{3},y_{4}\}.
\end{eqnarray*}
Finally, we obtain the following results:

$(1)$ $g$ is a homomorphism from $(U_{1}, \mathscr{R}^{\geq}_{A})$
to $(U_{2}, g(\mathscr{R}^{\geq}_{A}))$;

$(2)$ $g(R_{2})$, $g(R_{3})$ and $g(R_{4})$ are superfluous in
$g_{1}(\mathscr{R}^{\geq}_{A})$ if and only if $R_{2}$, $R_{3}$ and
$R_{4}$ are superfluous in $\mathscr{R}^{\geq}_{A}$;

$(3)$ $\{g(R_{1})\}$ is a reduct of $g(\mathscr{R}^{\geq}_{A})$ if
and only if $\{R_{1}\}$ is a reduct of $\mathscr{R}^{\geq}_{A}$.
\end{example}

\begin{table}[htbp]
\caption{The partitions based on $R_{1}, R_{2}, R_{3}, R_{4}$ and
$\mathscr{R}^{\geq}_{A}$, respectively.}
 \tabcolsep0.29in
\begin{tabular}{c c c c c c}
\hline $U_{1}$ &$R_{1}$& $R_{2}$ &$R_{3}$& $R_{4}$& $\mathscr{R}^{\geq}_{A}$\\
\hline
$x_{1}$ & $\{x_{1}, x_{7}\}$& $\{x_{1}, x_{7}\}$ & $U_{1}$&$U_{1}$ & $\{x_{1}, x_{7}\}$\\
$x_{2}$ & $\{x_{2}, x_{4}\}$& $\{x_{2},x_{3}, x_{4}\}$ &$U_{1}$&$U_{1}$& $\{x_{2}, x_{4}\}$\\
$x_{3}$ & $\{x_{3}, x_{5},x_{6},x_{8}\}$& $\{x_{2},x_{3}, x_{4}\}$ &$U_{1}$&$U_{1}$ & $\{x_{3}\}$\\
$x_{4}$ & $\{x_{2}, x_{4}\}$& $\{x_{2},x_{3}, x_{4}\}$ &$U_{1}$&$U_{1}$& $\{x_{2}, x_{4}\}$\\
$x_{5}$ & $\{x_{3}, x_{5},x_{6},x_{8}\}$& $\{x_{5}, x_{6}, x_{8}\}$& $U_{1}$&$U_{1}$ & $\{x_{5},x_{6}, x_{8}\}$\\
$x_{6}$ & $\{x_{3}, x_{5},x_{6},x_{8}\}$& $\{x_{5}, x_{6}, x_{8}\}$&$U_{1}$&$U_{1}$& $\{x_{5},x_{6}, x_{8}\}$\\
$x_{7}$ & $\{x_{1}, x_{7}\}$& $\{x_{1}, x_{7}\}$ &$U_{1}$ &$U_{1}$ & $\{x_{1}, x_{7}\}$\\
$x_{8}$ & $\{x_{3}, x_{5},x_{6},x_{8}\}$& $\{x_{5}, x_{6}, x_{8}\}$&$U_{1}$&$U_{1}$& $\{x_{5},x_{6}, x_{8}\}$\\
\hline
\end{tabular}
\end{table}

From Example 4.3, we see that the image system $(U_{2},
g(\mathscr{R}^{\geq}_{A}))$ has the relatively smaller size than the
original system $(U_{1}, \mathscr{R}^{\geq}_{A})$, and their
attribute reductions are equivalent to each other under the
condition of a homomorphism.

To illustrate that the time complexity of computing attribute
reductions is reduced greatly by means of homomorphisms from another
view, we employ an example to show attribute reductions on the basis
of the discernibility matrix in the following.

\begin{example} (Continuation of Example 4.3) Based on Definition
3.7, we obtain the discernibility matrixes $D_{1}$ and $D_{2}$ of
$(U_{1}, \mathscr{R}^{\geq}_{A})$ and $(U_{2},
g(\mathscr{R}^{\geq}_{A}))$, respectively.
$$
D_{1}=\left[
\begin{array}{ccccccc}
 \emptyset&    &    &    &   &    &    \\
\{a_{1}\} & \emptyset &    &   &    &    &     \\
\emptyset & \emptyset & \emptyset  &   &    &    &     \\
$\{$a_{1},a_{2}$\}$ & \emptyset & \emptyset& \emptyset  &    &    &    \\
$\{$a_{1},a_{2}$\}$ &\emptyset & \emptyset&\emptyset & \emptyset &    &    \\
\emptyset & \emptyset & \{a_{1}\}& \emptyset  & \{a_{1},a_{2}\} & \{a_{1},a_{2}\} &    \\
$ \{$a_{1},a_{2}$\}$ & \emptyset & \emptyset& \emptyset & \emptyset & \emptyset & \{a_{1},a_{2}\} \\
\end{array}
\right],
$$

and
$$D_{2}=\left[
\begin{array}{ccc}
\emptyset &    &     \\
\{a_{1}\} & \emptyset &    \\
\{a_{1},a_{2}\} & \emptyset & \emptyset \\
\end{array}
\right].
$$

It is obvious that the size of $D_{1}$ is larger than that of
$D_{2}$, and $\{a_{1}\}$ is the reduct of $(U_{1},
\mathscr{R}^{\geq}_{A})$ and $(U_{2}, g(\mathscr{R}^{\geq}_{A}))$.
We see that the time complexity of computing $D_{2}$ is relatively
lower than that of computing $D_{1}$.
\end{example}

From the practical viewpoint, it may be difficult to construct
attribute reducts of a large-scale set-valued information system
directly. However, we can convert it into a relation information
system and compress the relation information system into a
relatively smaller one under the condition of a homomorphism. Then
we conduct the attribute reductions of the image system which is
equivalent to that of the original information system. Therefore,
the homomorphisms may provide a more efficient approach to dealing
with attribute reductions of large-scale set-valued information
systems.

\section{Data compression of the dynamic set-valued information system}

In this section, we consider the data compression of four types of
dynamic set-valued information systems in terms of variations of the
attribute and object sets.

\subsection{Compressing the dynamic set-valued information system
when adding an attribute set}

Suppose $S_{1}=(U_{1}, A, V_{1}, f_{1})$ is a set-valued information
system. By adding an attribute set $P$ into $A$ satisfying $A\cap
P=\emptyset$, where $P=\{a_{m+1},a_{m+2},...,a_{k}\}$, we get the
updated set-valued information system $S_{2}=(U_{1}, A\cup P, V_{2},
f_{2})$. There are three steps to compress $S_{2}$ by utilizing the
compression of the original system $S_{1}$. First, we obtain the
induced $\geq-$relation information system $(U_{1},
\mathscr{R}^{\geq}_{P})$ and derive the partition $U_{1}/R_{i}$
based on $R_{i}\in \mathscr{R}^{\geq}_{P}$ $(m+1\leq i\leq k)$.
Second, we get Table 5 by adding the partition $U_{1}/R_{i}$
$(m+1\leq i\leq k)$ into Table 2 and derive the partition
$U_{1}/\mathscr{R}^{\geq}_{A\cup P}$. Third, as Example 4.3, we
define the homomorphism $g$ based on
$U_{1}/\mathscr{R}^{\geq}_{A\cup P}$ and derive the relation
information system $S_{3}=(g(U_{1}), g(\mathscr{R}^{\geq}_{A\cup
P}))$.

\begin{table}[htbp]
\caption{The partitions based on each relation $R_{i}$ $(1\leq i\leq
k)$ and $\mathscr{R}^{\geq}_{A\cup P}$, respectively.}
 \tabcolsep0.3in
\begin{tabular}{c c c c c c c c}
\hline $U_{1}$  &$R_{1}$& $R_{2}$&.&.&. &$R_{k}$& $\mathscr{R}^{\geq}_{A\cup P}$\\
\hline
$x_{1}$ & $P_{1x_{1}}$& $P_{2x_{1}}$&.&.&. &$P_{kx_{1}}  $ & $P_{(A\cup P) x_{1}}$\\
$x_{2}$ & $P_{1x_{2}}$& $P_{2x_{2}}$&.&.&. &$P_{kx_{2}}  $&$P_{(A\cup P) x_{2}}$\\
$.$     & $.$        & .&.&.&. &$ . $& .\\
$.$     & $.$        & .&.&.&. &$ . $&.\\
$.$     & $.$        & .&.&.&. &$ . $ &.\\
$x_{n}$ & $P_{1x_{n}}$& $P_{2x_{2}}$&.&.&. &$P_{kx_{n}}$&$P_{(A\cup P) x_{n}}$  \\
\hline
\end{tabular}
\end{table}

The following example is employed to illustrate the data compression
of dynamic set-valued information systems when adding an attribute
set.

\begin{table}[htbp]
\caption{A set-valued information system by adding an attribute
$a_{5}$ into Table 2.}
 \tabcolsep0.33in
\begin{tabular}{c c c c c c}
\hline $U_{1}$  & $a_{1}$ &$a_{2}$& $a_{3}$& $a_{4}$& $a_{5}$\\
\hline
$x_{1}$ & $\{0\}$& $\{0\}$ &$\{1,2\}$& $\{1,2\}$ & $\{1,2\}$\\
$x_{2}$ & $\{0,1,2\}$& $\{0,1,2\}$  &$\{1,2\}$& $\{0,1,2\}$&$\{0,2\}$\\
$x_{3}$ & $\{1,2\}$& $\{0,1\}$  &$\{1,2\}$& $\{1,2\}$& $\{1,2\}$\\
$x_{4}$ & $\{0,1\}$& $\{0,2\}$  &$\{1,2\}$& $\{1\}$& $\{2\}$\\
$x_{5}$ & $\{1,2\}$& $\{1,2\}$ &$\{1,2\}$  & $\{1\}$& $\{2\}$ \\
$x_{6}$ & $\{1\}$& $\{1,2\}$&$\{0,1\}$ & $\{0,1\}$& $\{0,1,2\}$\\
$x_{7}$ & $\{0\}$& $\{0\}$ &$\{1,2\}$& $\{1,2\}$ & $\{0,2\}$\\
$x_{8}$ & $\{1\}$& $\{1,2\}$&$\{0,1\}$ & $\{0,1\}$& $\{3\}$\\
\hline
\end{tabular}
\end{table}

\begin{example}
We obtain the updated set-valued information system shown in Table 6
by adding an attribute $a_{5}$ into the set-valued information
system shown in Table 2. By Definition 4.1, we first get that
$U_{1}/R_{5}=\{\{x_{1},x_{2},x_{3},x_{4},x_{5},x_{6},x_{7}\},$
$\{x_{8}\}\}$ based on $a_{5}$. Then we obtain Table 7 and derive
$U_{1}/\mathscr{R}^{\geq}_{A\cup
\{a_{5}\}}=\{\{x_{1},x_{7}\},\{x_{2},x_{4}\},\{x_{3}\},\{x_{5},x_{6}\},$
$\{x_{8}\}\}$. Afterwards, we define the mapping $g:
U_{1}\longrightarrow U_{2}$ as follows:
$$g(x_{1})=g(x_{7})= y_{1}, g(x_{2}) = g(x_{4}) = y_{2},
g(x_{3})= y_{3}, g(x_{5})=g(x_{6})=y_{4}, g(x_{8})= y_{5},$$ where
$U_{2}=\{y_{1},y_{2},y_{3},y_{4},y_{5}\}$. Consequently, we obtain
the relation information system $(U_{2}, g(\mathscr{R}^{\geq}_{A\cup
\{a_{5}\}}))$. For simplicity, we do not list the relation
information system in this subsection.

\begin{table}[htbp]
\caption{The partitions based on $R_{1}, R_{2}, R_{3}, R_{4}, R_{5}$
and $\mathscr{R}^{\geq}_{A\cup \{a_{5}\}}$, respectively.}
 \tabcolsep0.16in
\begin{tabular}{c c c c c c c}
\hline $U_{1}$ &$R_{1}$& $R_{2}$ &$R_{3}$& $R_{4}$&$R_{5}$& $\mathscr{R}^{\geq}_{A\cup \{a_{5}\}}$\\
\hline
$x_{1}$ & $\{x_{1}, x_{7}\}$& $\{x_{1}, x_{7}\}$ & $\{U_{1}\}$&$\{U_{1}\}$ &$\{x_{1}, x_{2}, x_{3}, x_{4}, x_{5}, x_{6}, x_{7}\}$& $\{x_{1}, x_{7}\}$\\
$x_{2}$ & $\{x_{2}, x_{4}\}$& $\{x_{2},x_{3}, x_{4}\}$ &$\{U_{1}\}$&$\{U_{1}\}$& $\{x_{1}, x_{2}, x_{3}, x_{4}, x_{5}, x_{6}, x_{7}\}$&$\{x_{2}, x_{4}\}$\\
$x_{3}$ & $\{x_{3}, x_{5},x_{6},x_{8}\}$& $\{x_{2},x_{3}, x_{4}\}$ &$\{U_{1}\}$&$\{U_{1}\}$ & $\{x_{1}, x_{2}, x_{3}, x_{4}, x_{5}, x_{6}, x_{7}\}$&$\{x_{3}\}$\\
$x_{4}$ & $\{x_{2}, x_{4}\}$& $\{x_{2},x_{3}, x_{4}\}$ &$\{U_{1}\}$&$\{U_{1}\}$& $\{x_{1}, x_{2}, x_{3}, x_{4}, x_{5}, x_{6}, x_{7}\}$&$\{x_{2}, x_{4}\}$\\
$x_{5}$ & $\{x_{3}, x_{5},x_{6},x_{8}\}$& $\{x_{5}, x_{6}, x_{8}\}$& $\{U_{1}\}$&$\{U_{1}\}$ & $\{x_{1}, x_{2}, x_{3}, x_{4}, x_{5}, x_{6}, x_{7}\}$&$\{x_{5},x_{6}\}$\\
$x_{6}$ & $\{x_{3}, x_{5},x_{6},x_{8}\}$& $\{x_{5}, x_{6}, x_{8}\}$&$\{U_{1}\}$&$\{U_{1}\}$& $\{x_{1}, x_{2}, x_{3}, x_{4}, x_{5}, x_{6}, x_{7}\}$&$\{x_{5},x_{6}\}$\\
$x_{7}$ & $\{x_{1}, x_{7}\}$& $\{x_{1}, x_{7}\}$ &$\{U_{1}\}$ &$\{U_{1}\}$ & $\{x_{1}, x_{2}, x_{3}, x_{4}, x_{5}, x_{6}, x_{7}\}$&$\{x_{1}, x_{7}\}$\\
$x_{8}$ & $\{x_{3}, x_{5},x_{6},x_{8}\}$& $\{x_{5}, x_{6}, x_{8}\}$&$\{U_{1}\}$&$\{U_{1}\}$&$\{x_{8}\}$& $\{x_{8}\}$\\
\hline
\end{tabular}
\end{table}
\end{example}

In Example 5.1, we compress the dynamic set-valued information
system when adding an attribute. The same approach can be applied to
the dynamic set-valued information system when adding an attribute
set.

\subsection{Compressing the dynamic set-valued information system
when deleting an attribute set}

Suppose $S_{1}=(U_{1}, A, V_{1}, f_{1})$ is a set-valued information
system. By deleting an attribute $a_{l}\in A$, we get the updated
set-valued information system $S_{2}=(U_{1}, A-\{a_{l}\}, V_{2},
f_{2})$. First, we obtain Table 8 by deleting the partition
$U_{1}/R_{l}$ shown in Table 2. Second, we get the partition
$U/\mathscr{R}^{\geq}_{(A-\{a_{l}\})}$ based on $U_{1}/R_{i}$
$(1\leq i\leq l-1, l+1\leq i\leq m )$ and define the homomorphism
$g$ as Example 4.3. Third, we obtain the relation information system
$S_{3}=(g(U_{1}), g(\mathscr{R}^{\geq}_{(A-\{a_{l}\})}))$. We can
compress the dynamic set-valued information system when deleting an
attribute set with the same approach.
\begin{table}[htbp]
\caption{The partitions based on each covering $R_{i}$ $(1\leq i\leq
l-1, l+1\leq i\leq m )$ and $\mathscr{R}^{\geq}_{(A-\{a_{l}\})}$,
respectively.}
 \tabcolsep0.145in
\begin{tabular}{c c c c c c c c c c c c c}
\hline $U_{1}$  &$R_{1}$&$R_{2}$&.&.&.&$R_{l-1}$&$R_{l+1}$&.&.&.&$R_{m}$& $\mathscr{R}^{\geq}_{(A-\{a_{l}\})}$\\
\hline
$x_{1}$ & $P_{1x_{1}}$&$P_{2x_{1}}$&.&.&.&$P_{(l-1)x_{1}}$&$P_{(l+1)x_{1}}$&.&.&. &$P_{mx_{1}}$& $P_{(A-\{a_{l}\}) x_{1}}$\\
$x_{2}$ & $P_{1x_{2}}$&$P_{2x_{2}}$&.&.&.&$P_{(l-1)x_{2}}$&$P_{(l+1)x_{2}}$&.&.&. &$P_{mx_{2}}$&$P_{(A-\{a_{l}\}) x_{2}}$\\
$.$     & $.$        & .&.&.&.&.&.&.&.&. &$ . $& .\\
$.$     & $.$        & .&.&.&.&.&.&.&.&. &$ . $& .\\
$.$     & $.$        & .&.&.&.&.&.&.&.&. &$ . $& .\\
$x_{n}$ & $P_{1x_{n}}$&$P_{2x_{n}}$&.&.&.&$P_{(l-1)x_{n}}$&$P_{(l+1)x_{n}}$&.&.&. &$P_{mx_{n}}$&$P_{(A-\{a_{l}\}) x_{n}}$  \\
\hline
\end{tabular}
\end{table}

We employ an example to illustrate that how to compress the dynamic
set-valued information system when deleting an attribute set as
follows.

\begin{table}[htbp]
\caption{A set-valued information system.}
 \tabcolsep0.62in
\begin{tabular}{ c c c c}
\hline $U_{1}$  & $a_{2}$& $a_{3}$& $a_{4}$\\ \hline
$x_{1}$ & $\{0\}$ &$\{1,2\}$& $\{1,2\}$ \\
$x_{2}$ & $\{0,1,2\}$  &$\{1,2\}$& $\{0,1,2\}$\\
$x_{3}$ & $\{0,1\}$  &$\{1,2\}$& $\{1,2\}$\\
$x_{4}$ & $\{0,2\}$  &$\{1,2\}$& $\{1\}$\\
$x_{5}$ & $\{1,2\}$ &$\{1,2\}$  & $\{1\}$ \\
$x_{6}$ & $\{1,2\}$&$\{0,1\}$ & $\{0,1\}$\\
$x_{7}$ & $\{0\}$ &$\{1,2\}$& $\{1,2\}$ \\
$x_{8}$ & $\{1,2\}$&$\{0,1\}$ & $\{0,1\}$\\
\hline
\end{tabular}
\end{table}

\begin{table}[htbp]
\caption{The partitions based on $ R_{2}, R_{3}, R_{4}$ and
$\mathscr{R}^{\geq}_{(A-\{a_{1}\})}$, respectively.}
 \tabcolsep0.45in
\begin{tabular}{ c c c c c}
\hline $U_{1}$ & $R_{2}$ &$R_{3}$& $R_{4}$& $\mathscr{R}^{\geq}_{(A-\{a_{1}\})}$\\
\hline
$x_{1}$& $\{x_{1}, x_{7}\}$ & $\{U_{1}\}$&$\{U_{1}\}$ & $\{x_{1}, x_{7}\}$\\
$x_{2}$& $\{x_{2},x_{3}, x_{4}\}$ &$\{U_{1}\}$&$\{U_{1}\}$& $\{x_{2}, x_{3},x_{4}\}$\\
$x_{3}$& $\{x_{2},x_{3}, x_{4}\}$ &$\{U_{1}\}$&$\{U_{1}\}$ & $\{x_{2}, x_{3},x_{4}\}$\\
$x_{4}$& $\{x_{2},x_{3}, x_{4}\}$ &$\{U_{1}\}$&$\{U_{1}\}$& $\{x_{2}, x_{3},x_{4}\}$\\
$x_{5}$& $\{x_{5}, x_{6}, x_{8}\}$& $\{U_{1}\}$&$\{U_{1}\}$ & $\{x_{5},x_{6},x_{8}\}$\\
$x_{6}$& $\{x_{5}, x_{6}, x_{8}\}$&$\{U_{1}\}$&$\{U_{1}\}$& $\{x_{5},x_{6},x_{8}\}$\\
$x_{7}$& $\{x_{1}, x_{7}\}$ &$\{U_{1}\}$ &$\{U_{1}\}$ & $\{x_{1}, x_{7}\}$\\
$x_{8}$& $\{x_{5}, x_{6}, x_{8}\}$&$\{U_{1}\}$&$\{U_{1}\}$& $\{x_{5},x_{6},x_{8}\}$\\
\hline
\end{tabular}
\end{table}

\begin{example}
By deleting the attribute $a_{1}$ in the set-valued information
system $S_{1}$ shown in Table 3, we obtain the updated set-valued
information system $S_{2}$ shown in Table 9. To compress the updated
information system $S_{2}$ based on the compression of $S_{1}$, we
get Table 10 by deleting $U_{1}/R_{1}$ based on $a_{1}$.  Then we
obtain the partition
$U_{1}/\mathscr{R}^{\geq}_{(A-\{a_{1}\})}=\{\{x_{1},
x_{7}\},\{x_{2}, x_{3}, x_{4}\},\{x_{5}, x_{6}, x_{8}\}\}$ and
define the mapping $g: U_{1}\longrightarrow U_{2}$ as follows:
$$g(x_{1})=g(x_{7})= y_{1}, g(x_{2})=g(x_{3})
=g(x_{4}) = y_{2}, g(x_{5})=g(x_{6})=g(x_{8})=y_{3},$$ where
$U_{2}=\{y_{1},y_{2},y_{3}\}$. Subsequently, the set-valued
information system $(U_{1}, A-\{a_{1}\}, V, f_{1})$ can be
compressed into a relatively small relation system $(U_{2},
\{g(R_{2}),g(R_{3}),g(R_{4})\})$. To express clearly, we do not list
all the relations in this subsection.
\end{example}

In Example 5.2, we compress the dynamic set-valued information
system when deleting an attribute. The same approach can be applied
to the set-valued information system when deleting an attribute set.

\subsection{Compressing the dynamic set-valued information system
when adding an object set}

In this subsection, we introduce the equivalence relation for the
set-valued information system.

\begin{definition}
Let $S_{1}=(U_{1}, A, V, f_{1})$ be a set-valued information system.
Then the equivalence relation $T_{A}$ is defined as
$$T_{A}=\{(x, y)|\forall a\in A, f(x,a)= f(y,a), x, y\in U_{1}\}.$$
\end{definition}

It is obvious that Pawlak's equivalence relation is the same as that
given in Definition 5.3 if the set-valued information system is
classical. For the sake of convenience, we denote
$[x]^{1}_{A}=\{y|(x,y)\in T_{A},  x,y\in U_{1}\}$. There are two
steps to compress $S_{1}=(U_{1}, A, V, f_{1})$ based on $T_{A}$. We
first derive the partition $U_{1}/A=\{C_{1}, C_{2},...,C_{N}\}$ on
the basis of $T_{A}$. Then we define $g_{1}(x)=y_{k}$ for any $x\in
C_{k}$ and obtain $S_{2}=(U_{2}, A, V, f_{2})$, where
$U_{2}=\{y_{k}|1\leq k\leq N\}, f_{2}(y_{k}, a)=f_{1}(x,a)$ for
$a\in A$, and $x\in g_{1}^{-1}(y_{k})$. Suppose we obtain
$S_{4}=(U_{1}\cup U_{3}, A, V, f_{1}\cup f_{2})$ by adding the
set-valued information system $S_{3}=(U_{3}, A, V, f_{3})$ into
$S_{1}$. To compress $S_{4}$ by utilizing the compression of the
original system $S_{1}$, first, we obtain $S_{5}$ by compressing
$S_{3}$ as $S_{1}$. Second, we compress $S_{2}\cup S_{5}$ as $S_{1}$
and get $S_{7}$ which is the same as the compression of $S_{1}\cup
S_{3}$. To express clearly, the process of the compression of
set-valued information systems can be illustrated as follows:

\begin{equation*} \label{eq:1}
\begin{aligned}
         S_{1}&\looparrowright  S_{2} \\
         S_{3}&\looparrowright  S_{5}
         \end{aligned} \left\}S_{6}=S_{2}\cup S_{5}\looparrowright S_{7}
         \looparrowleft
\begin{aligned}
          &  \\
          &
          \end{aligned} S_{4}=S_{1}\cup S_{3}\right\{\begin{aligned}
          S_{1}&  \\
          S_{3}&
          \end{aligned},
          \end{equation*}
where $\looparrowright$ (respectively, $\looparrowleft$) denotes the
process of the compression of set-valued information systems.

\begin{table}[htbp]
\caption{The set-valued information system $S_{1}$.}
 \tabcolsep0.65in
\begin{tabular}{c c c c }
\hline $U_{1}$  & $a_{1}$ &$a_{2}$& $a_{3}$\\ \hline
$x_{1}$ & $\{0,1\}$& $\{0,2\}$ &$\{1,2\}$\\
$x_{2}$ & $\{0,1\}$& $\{0,2\}$  &$\{1,2\}$\\
$x_{3}$ & $\{0,1\}$& $\{1\}$  &$\{0,1\}$\\
$x_{4}$ & $\{0,1\}$& $\{1\}$  &$\{0,1\}$\\
$x_{5}$ & $\{1,2\}$& $\{1\}$ &$\{1,2\}$ \\
$x_{6}$ & $\{1,2\}$& $\{1\}$&$\{1,2\}$ \\
\hline
\end{tabular}
\end{table}

We employ an example to illustrate the data compression of
set-valued information systems.

\begin{example}
Table 11 shows the set-valued information system $S_{1}=\{U_{1}, A,
V, f_{1}\}$. By Definition 5.3, we obtain that $U_{1}/A=\{\{x_{1},
x_{2}\},\{x_{3}, x_{4}\},\{x_{5}, x_{6}\}\}$. Then we define $g_{1}$
and $f_{2}$ as follows:
$$g_{1}(x_{1})=g_{1}(x_{2})=y_{1},
g_{1}(x_{3})=g_{1}(x_{4})=y_{2}, g_{1}(x_{5})=g_{1}(x_{6})=y_{3},
f_{2}(y_{i}, a_{i})=f_{1}(x, a_{i}),$$ where $x\in
g_{1}^{-1}(y_{i})$. Thus we can compress $S_{1}$ into
$S_{2}=(U_{2},A, V,f_{2})$, where $U_{2}=\{g(x)|x\in U_{1}\}$, and
$S_{2}$ is shown in Table 12.

\begin{table}[htbp]
\caption{The compressed set-valued information system $S_{2}$ of
$S_{1}$.}
 \tabcolsep0.65in
\begin{tabular}{c c c c }
\hline $U_{2}$  & $a_{1}$ &$a_{2}$& $a_{3}$\\ \hline
$y_{1}$ & $\{0,1\}$& $\{0,2\}$ &$\{1,2\}$\\
$y_{2}$ & $\{0,1\}$& $\{1\}$  &$\{0,1\}$\\
$y_{3}$ & $\{1,2\}$& $\{1\}$ &$\{1,2\}$ \\
\hline
\end{tabular}
\end{table}
\end{example}

\begin{table}[htbp]
\caption{The set-valued information system $S_{3}$.}
 \tabcolsep0.65in
\begin{tabular}{c c c c }
\hline $U_{3}$  & $a_{1}$ &$a_{2}$& $a_{3}$\\ \hline
$x_{7}$ & $\{1,2\}$& $\{0,2\}$ &$\{0,1\}$ \\
$x_{8}$ & $\{1,2\}$& $\{0,2\}$&$\{0,1\}$ \\
$x_{9}$ & $\{0,1\}$& $\{1\}$  &$\{0,1\}$\\
$x_{10}$ & $\{0,1\}$& $\{1\}$  &$\{0,1\}$\\
\hline
\end{tabular}
\end{table}

\begin{table}[htbp]
\caption{The set-valued information system $S_{4}=S_{1}\cup S_{3}$.}
 \tabcolsep0.57in
\begin{tabular}{c c c c }
\hline $U_{4}=U_{1}\cup U_{3}$  & $a_{1}$ &$a_{2}$& $a_{3}$\\ \hline
$x_{1}$ & $\{0,1\}$& $\{0,2\}$ &$\{1,2\}$\\
$x_{2}$ & $\{0,1\}$& $\{0,2\}$  &$\{1,2\}$\\
$x_{3}$ & $\{0,1\}$& $\{1\}$  &$\{0,1\}$\\
$x_{4}$ & $\{0,1\}$& $\{1\}$  &$\{0,1\}$\\
$x_{5}$ & $\{1,2\}$& $\{1\}$ &$\{1,2\}$ \\
$x_{6}$ & $\{1,2\}$& $\{1\}$&$\{1,2\}$ \\
$x_{7}$ & $\{1,2\}$& $\{0,2\}$ &$\{0,1\}$ \\
$x_{8}$ & $\{1,2\}$& $\{0,2\}$&$\{0,1\}$ \\
$x_{9}$ & $\{0,1\}$& $\{1\}$  &$\{0,1\}$\\
$x_{10}$ & $\{0,1\}$& $\{1\}$  &$\{0,1\}$\\
\hline
\end{tabular}
\end{table}

\begin{table}[htbp]
\caption{The set-valued information system $S_{5}$.}
 \tabcolsep0.65in
\begin{tabular}{c c c c }
\hline $U_{5}$  & $a_{1}$ &$a_{2}$& $a_{3}$\\ \hline
$y_{4}$ & $\{1,2\}$& $\{0,2\}$ &$\{0,1\}$ \\
$y_{5}$ & $\{0,1\}$& $\{1\}$  &$\{0,1\}$\\
\hline
\end{tabular}
\end{table}

\begin{table}[htbp]
\caption{The set-valued information system $S_{6}=S_{2}\cup S_{5}$.}
 \tabcolsep0.60in
\begin{tabular}{c c c c }
\hline $U_{6}=U_{2}\cup U_{4} $  & $a_{1}$ &$a_{2}$& $a_{3}$\\
\hline
$y_{1}$ & $\{0,1\}$& $\{0,2\}$ &$\{1,2\}$\\
$y_{2}$ & $\{0,1\}$& $\{1\}$  &$\{0,1\}$\\
$y_{3}$ & $\{1,2\}$& $\{1\}$ &$\{1,2\}$ \\
$y_{4}$ & $\{1,2\}$& $\{0,2\}$ &$\{0,1\}$ \\
$y_{5}$ & $\{0,1\}$& $\{1\}$  &$\{0,1\}$\\
\hline
\end{tabular}
\end{table}

\begin{table}[htbp]
\caption{The set-valued information system $S_{7}$.}
 \tabcolsep0.65in
\begin{tabular}{c c c c }
\hline $U_{7}$  & $a_{1}$ &$a_{2}$& $a_{3}$\\ \hline
$z_{1}$ & $\{0,1\}$& $\{0,2\}$ &$\{1,2\}$\\
$z_{2}$ & $\{0,1\}$& $\{1\}$  &$\{0,1\}$\\
$z_{3}$ & $\{1,2\}$& $\{1\}$ &$\{1,2\}$ \\
$z_{4}$ & $\{1,2\}$& $\{0,2\}$ &$\{0,1\}$ \\
\hline
\end{tabular}
\end{table}

The following example is employed to illustrate how to update the
compression when adding an object set.

\begin{example}
By adding $S_{3}$ shown in Table 13 into $S_{1}$, we obtain the
set-valued information system $S_{4}=S_{1}\cup S_{3}$ shown in Table
14. To compress $S_{4}$, as Example 5.4, we compress $S_{3}$ to
$S_{5}=(U_{5}, A, V, f_{5})$ shown in Table 15. Then we compress
$S_{6}=S_{2}\cup S_{5}$ shown in Table 16 and obtain $S_{7}=\{U_{7},
A, V, f_{7}\}$ shown in Table 17. Afterwards, we can continue to
compress $S_{7}$ as Example 4.3 in Section 4.
\end{example}

\subsection{Compressing the dynamic set-valued information systems
when deleting an object set}

Suppose $S_{1}=(U_{1}, A, V, f_{1})$ is a set-valued information
system, we compress $S_{1}$ to $S_{2}=(U_{2}, A, V, f_{2})$ under
the condition of a homomorphism $g_{1}$. By deleting $S_{3}=(U_{3},
A, V, f_{3})$, we obtain $S_{4}=(U_{4}, A, V, f_{4})$, where
$U_{3}\subseteq U_{1}$ and $U_{4}=U_{1}-U_{3}$. There are three
steps to compress $S_{4}=(U_{4}, A, V, f_{4})$ based on $S_{2}$. By
Definition 5.3, we first obtain that $U_{1}/A=\{[x]^{1}_{A}|x\in
U_{1}\}$ and $U_{3}/A=\{[x]^{3}_{A}|x\in U_{3}\}$. It is obvious
that $[x]^{3}_{A}\subseteq [x]^{1}_{A}$ for any $x\in U_{3}$. Then
we cancel the object $g_{1}(x)$ in $U_{2}$ if
$[x]^{3}_{A}=[x]^{1}_{A}$ and keep the object $g_{1}(x)$ in $U_{2}$
if $[x]^{3}_{A}\neq[x]^{1}_{A}$. Third, we obtain the set-valued
information system $S_{5}=(U_{5}, A, V, f_{5})$ after the deletion.

Following, we employ an example to illustrate the process of the
compression of the updated set-valued information system.

\begin{table}[htbp]
\caption{The set-valued information system $S_{3}$.}
 \tabcolsep0.65in
\begin{tabular}{c c c c }
\hline $U_{3}$  & $a_{1}$ &$a_{2}$& $a_{3}$\\ \hline
$x_{1}$ & $\{0,1\}$& $\{0,2\}$ &$\{1,2\}$\\
$x_{2}$ & $\{0,1\}$& $\{0,2\}$  &$\{1,2\}$\\
$x_{3}$ & $\{0,1\}$& $\{1\}$  &$\{0,1\}$\\
\hline
\end{tabular}
\end{table}

\begin{table}[htbp]
\caption{The set-valued information system $S_{4}$.}
 \tabcolsep0.57in
\begin{tabular}{c c c c }
\hline $U_{4}=U_{1}-U_{3}$  & $a_{1}$ &$a_{2}$& $a_{3}$\\ \hline
$x_{4}$ & $\{0,1\}$& $\{1\}$  &$\{0,1\}$\\
$x_{5}$ & $\{1,2\}$& $\{1\}$ &$\{1,2\}$ \\
$x_{6}$ & $\{1,2\}$& $\{1\}$&$\{1,2\}$ \\
$x_{7}$ & $\{1,2\}$& $\{0,2\}$ &$\{0,1\}$ \\
$x_{8}$ & $\{1,2\}$& $\{0,2\}$&$\{0,1\}$ \\
$x_{9}$ & $\{0,1\}$& $\{1\}$  &$\{0,1\}$\\
$x_{10}$ & $\{0,1\}$& $\{1\}$  &$\{0,1\}$\\
\hline
\end{tabular}
\end{table}

\begin{example} We take information systems $S_{4}$ and
$S_{7}$ in Example 5.5 as the original set-valued information system
$S_{1}$ and the compression information system $S_{2}$,
respectively. By deleting $S_{3}=(U_{3}, A, V, f)$ shown in Table
18, where $U_{3}=\{x_{1},x_{2},x_{3}\}$, we obtain the set-valued
information system $S_{4}$ shown in Table 19. To compress $S_{4}$,
we first get that
$U_{1}/A=\{\{x_{1},x_{2}\},\{x_{3},x_{4},x_{9},x_{10}\}, $
$\{x_{5},x_{6}\},\{x_{7},x_{8}\}\}$ and
$U_{3}/A=\{\{x_{1},x_{2}\},\{x_{3}\}\}$. Obviously,
$[x_{1}]^{1}_{A}=[x_{2}]^{1}_{A}=\{x_{1},
x_{2}\}=[x_{1}]^{3}_{A}=[x_{2}]^{3}_{A}$ and
$[x_{3}]^{3}_{A}=\{x_{3}\}\subset\{x_{3}, x_{4},x_{9},
x_{10}\}=[x_{3}]^{1}_{A}$. Then we cancel $z_{1}$ and keep $\{z_{2},
z_{3}, z_{4}\}$ in Table 17. Afterwards, we obtain the compressed
set-valued information system $S_{5}$ shown in Table 20. We can
continue to compress $S_{5}$ as Example 4.3 in Section 4.

\begin{table}[htbp]
\caption{The set-valued information system $S_{5}$.}
 \tabcolsep0.65in
\begin{tabular}{c c c c }
\hline $U_{5}$  & $a_{1}$ &$a_{2}$& $a_{3}$\\ \hline
$z_{2}$ & $\{0,1\}$& $\{1\}$  &$\{0,1\}$\\
$z_{3}$ & $\{1,2\}$& $\{1\}$ &$\{1,2\}$ \\
$z_{4}$ & $\{1,2\}$& $\{0,2\}$ &$\{0,1\}$ \\
\hline
\end{tabular}
\end{table}
\end{example}

\section{Conclusions}

In this paper, we have proposed three tolerance relations for the
set-valued information system and studied their basic properties.
Then the data compression of set-valued information systems has been
discussed in detail. Afterwards, we have studied the data
compression of dynamic set-valued information systems by using the
precious compression of the original set-valued information systems.

In the future, we will study the data compression of fuzzy
set-valued information systems and dynamic fuzzy set-valued
information systems. We will investigate the data compression of
interval-valued information systems, fuzzy interval-valued
information systems, dynamic interval-valued information systems and
dynamic fuzzy interval-valued information systems.

\section*{ Acknowledgments}

We would like to thank the anonymous reviewers very much for their
professional comments and valuable suggestions. This work is
supported by the National Natural Science Foundation of China (NO.
11071061) and the National Basic Research Program of China (NO.
2010CB334706, 2011CB311808).

\end{document}